\documentstyle{article}
\begin{document}
%\rightline{IFUG-95-R-3}
%\rightline{June 28, 1995}
\rightline{\scriptsize{cond-mat/9510019}}
\rightline{\scriptsize{Nuovo Cimento D 18, 477-481 (April 1996)}}
\begin{center}{\Large{\bf On the kinks and dynamical phase transitions of
$\alpha$-helix protein chains}\\

H. Rosu
\footnote{e-mail: rosu@ifug.ugto.mx}\\[2mm]
%\begin{titlepage}
%\title{ \bf xxxxxxxxxx}
%\author{Haret C. Rosu \\
%\\
{\scriptsize{Instituto de F\'{\i}sica, Universidad de Guanajuato,\\
Apdo Postal E-143, 37150 Le\'on, Gto, M\'exico\\  }}}
\end{center}

%\date{June 28, 1995}

%{\baselineskip=20pt
\begin{center}
{\bf Abstract}

Heuristic insights into a physical picture of
Davydov's solitonic model of the one-dimensional protein chain are presented
supporting
the idea of a non-equilibrium competition between the Davydov phase and
a complementary, dynamical- `ferroelectric' phase along the chain.

\vskip 0.5cm


\vskip 0.2cm


\end{center}

%\vskip 2cm

%PACS numbers: xxxxxxxxxxxx

%IFUG-95-R-3 [$\cal H \cal C \cal R$]

%\vskip 1.5cm

             %%%%%%%%%     THE PAPER      %%%%%%%%%%%%
                  %%%% written by H. Rosu %%%%%
                        %%% JUNE 1995 %%%

%%%%%%%%%%%%%%%%%%%%%%%%%%%%%%%%%%%%%%%%%%%%%%%%%%%%%%%%%%%%%%%%%%%%%%%%%%%%
%\section{Introduction}
%%%%%%%%%%%%%%%%%%%%%%%%%%%%%%%%%%%%%%%%%%%%%%%%%%%%%%%%%%%%%%%%%%%%%%%%%%%%

A well-known application of the cubic nonlinear Schr\"{o}dinger
equation (NLSE) in biology is Davydov's model of energy transport by
means of solitons along $\alpha$-helix protein chains
\cite{Dav85}.
Over the years, the
model was turned almost into a paradigm of bioenergetics \cite{s93},
although the experimental results are debatable \cite{Fan90}.

%%%%%%%%%%%%%%%%%%%%%%%%%%%%%%%%%%%%%%%%%%%%%%%%%%%%%%%%%%%%%%%%%%%%%%%%%%%%
%\section{ Davydov Model}
%%%%%%%%%%%%%%%%%%%%%%%%%%%%%%%%%%%%%%%%%%%%%%%%%%%%%%%%%%%%%%%%%%%%%%%%%%%%

The $\alpha$-helix is the most
common secondary structure of proteins entailing three spines
 in the longitudinal z-direction that we consider infinite in
extent and having the peptide sequence $(---H-N-C=O)_{n}$, $n=\infty$,
where the dashed lines represent hydrogen bonds. There are side
radicals and their order is characteristic to each protein. The
3 spines are weaved together in a sort of incommensurate quasi-one
dimensional structure, but here we shall consider only the simple
one-spine chain, containing the peptide groups as molecular units.

In the Davydov model the main assumption is that about half
of the chemical energy ( $\epsilon =0.422\;eV$) released in the
hydrolysis of adenosine
triphosphate (ATP) turns into the vibrational energy
($\epsilon _{0}=0.205\;eV$) of the self-trapped amide-I
(C=O stretching) mode of the peptide unit. The amplitude of the
amide mode is self-trapped in the form of a sech- envelope soliton
whenever there is a balance between
the dipolar nearest-neighbour interaction and an admittedly
quite strong nonlinear amide-phonon
interaction. The energy transfer in the standard Davydov model is through
the nonradiative resonant dipole- dipole interaction and a rather
ambiguous vibrational-acoustic coherent state, the so-called $D_{1}$
{\em ansatz}.
It is a mixture of quantum and classical Hamiltonian methods that has
been deeply scrutinized in the literature \cite{Blw86}. Davydov
Hamiltonian can be written in the form
%%%%%%%%%%%%%%%%%%%%%%%%%%%%%%%%%%%%%%%%%%%%%%%%%%%%%%%%%%
\begin{eqnarray}
    H_{D} & = & H_{C=O}+H_{ph}+H_{int}\; \; \label{eq:d1}\\
\;\;H_{C=O} & = & \sum_{n} {\epsilon _{0}B_{n}^{\dagger}B_{n}-
J(B_{n}^{\dagger}B_{n+1}+ B_{n+1}^{\dagger}B_{n})} \;\; \label{eq:d2}\\
   H_{ph} & = & \sum_{q}\hbar\Omega_{q}(b_{q}^{\dagger}b_{q}+\frac{1}{2})
\; \; \label{eq:d3}\\
H_{int} & = & \frac{1}{\sqrt{N}}
\sum_{q,n}\chi(q)e^{iqnR}B_{n}^{\dagger}B_{n}(b_{q}+b_{-q}^{\dagger})
\; \; \label{eq:d4}
\end{eqnarray}
%%%%%%%%%%%%%%%%%%%%%%%%%%%%%%%%%%%%%%%%%%%%%%%%%%%%%%%%%%%%
The capital and small operators are vibrational and phonon ones, respectively.
Brown \cite{Br88} has shown in a clear way that Davydov Hamiltonian is
a particular case of the general Fr\"ohlich Hamiltonian of polaronic
systems \cite{Fro54}
%%%%%%%%%%%%%%%%%%%%%%%%%%%%%%%%%%%%%%%%%%%%
\begin{equation}
H_{FP}=\sum _{mn}J_{mn}a_{m}^{\dagger}a_{n}+
\sum _{q} \hbar \omega _{q}b_{q}^{\dagger}b_{q}+
\sum _{qn} \hbar \omega _{q}(\chi _{n}^{q}b_{q}^{\dagger}
+\chi_{n}^{q*}b_{q})a_{n}^{\dagger}a_{n}   \label{eq:r3}
\end{equation}
%%%%%%%%%%%%%%%%%%%%%%%%%%%%%%%%%%%%%%%%%%%%%
with $D_{1}$ states satisfying the Schr\"odinger equation of the
Fr\"ohlich Hamiltonian in the limit $J_{mn}=0$ and another $D_{2}$
{\em ansatz} valid for Schr\"odinger evolution in the limit
$\chi _{r}^{q}=0$.

The NLS subsonic soliton solutions of the energy transport coming
out from the Davydov model are
%%%%%%%%%%%%%%%%%%%%%%%%%%%%%%%%%%%%%%%%

\begin{eqnarray}
\alpha(\xi) & = & \sqrt{\mu /2}\;
e^{[\frac{i}{\hbar}[\frac{\hbar ^{2}v_{s}x}{2JR^{2}}-E_{s}t]]}
\cosh ^{-1}(\frac{\mu}{R}\xi) \; \; \label{eq:dd1} \\
\rho (\xi) & = &\frac{\beta _0}{2}
sech ^{-2}(\frac{\mu}{R}\xi) \; \; \label{eq:dd2} \\
\beta (\xi) & = & \frac{\beta _0}{2}(1-\tanh(Q\xi))
\; \; \label{eq:dd3}
\end{eqnarray}
%%%%%%%%%%%%%%%%%%%%%%%%%%%%%%%%
where $\xi =x-v_{s}t$ is the moving frame coordinate, $J$ is
the hopping (dipole-dipole) constant, $\chi$ is the nonlinear
dipole-phonon coupling parameter, $\gamma _{s}=1/\sqrt{1-s^{2}}$
($s=v_{s}/v_{a}$) is the soliton
`relativistic' factor, $w$ is the elasticity constant of the chain,
$\mu =\chi ^{2} \gamma _{s}^{2}/Jw$,
$\beta _0 =\frac{2\chi \gamma _{s}^{2}}{w}$,
$Q=MR\chi ^{2}\gamma _{s}^{2}/2w\hbar ^{2}$, $E_{s}=\epsilon _{0}-2J
+\hbar ^{2} v_{s}^{2}/4JR^{2}- J\mu ^{2}/3$, $M$ is the peptide molecular
mass, and $R$ is the hydrogen-bond
length.
The first soliton is the vibrational one in which one may
contemplate the soliton energy $E_{s}$ self-trapped by the carrier wave.
The second solution is related to the local deformation produced by the
vibrational soliton in the lattice.
The $\beta$-kink is a domain-wall configuration of the displacements
of the peptide groups into the new equilibrium positions. It has an
enhanced stability of topological origin since all the peptide groups
from the right side of the kink ($\xi > 0$ ) are in nondisplaced
positions
whereas all the peptide groups at the left ($\xi < 0$) are displaced
by the same amount $\beta _{0}=\frac{2\chi \gamma _{s}^{2}}{w}$. In
order to destroy the domain wall configuration, one should first turn the
left peptide groups to their initial position. Now let us state the essential
point in our phenomenology which is to think of
{\it the slowly moving Davydov $\beta$-kink as an interphase boundary for the
non-equilibrium transition from the Davydov dynamical regime of the
polypeptide chain to a dynamic- `ferroelectric' phase of the chain}.
This is not the only possible interpretation.
Indeed, in the same `non-relativistic'
limit, the $\beta$-kink, {\em per se}, is the solution of a classical
anharmonic oscillator equation with a frictional term, but only for a
particular value of the dissipation coefficient \cite{Dix}. However,
here we shall pursue the first interpretation.
In the literature on ferroelectricity it is common to consider the
interfacial boundary as the kink solution of a time-dependent Ginzburg-Landau
(GL) equation \cite{Gor83}. Our interpretation is based on the fact that
for low subsonic regime ($s^{2}\ll 1$) the Davydov $\beta$-kink is just the
complement of the (quasi)-static double-well GL kink
(i.e., $K_{D}\propto (1-K_{GL})$).
Since we
are in a non-equilibrium situation the more precise terminology for these
kinks is dynamic
interphase boundaries or interfacial patterns. One would like to
study their morphology during the growth. This is a difficult task
since we are in a more complicated case as compared to the simple
solid on solid model, or the kinetic Ising model, where the
width of the mean field interface is given in terms of the nearest neighbor
exchange interaction \cite{Cho88}, corresponding to the $J$ parameter
in the Davydov model. The width $W=1/2Q$ of the $\beta$-kink is determined
by a number of parameters, among which the $J$ of nearest neighbors and
the $\chi$ parameter of
the nonlinear interaction are controling the interfacial morphology. It is also
known that in the mean field theory the intrinsic width is identical to the
correlation length in the direction perpendicular to the interface.

Perhaps, one should also notice the formal analogy between the form of the
Davydov kink and the Glauber transition rate in one-dimensional
spin chains. The message of this analogy is that the kink is just
a step structural function which is required by a Hamiltonian
evolution and by a detailed balance condition in the spatial coordinate.

%%%%%%%%%%%%%%%%%%%%%%%%%%%%%%%%%%%%%%%%%%%%%%%%%%%%%%%%%%%%%%%%%%%%%%%%%%%%%
%\section{Multifractality of $\beta$- kinks}
%%%%%%%%%%%%%%%%%%%%%%%%%%%%%%%%%%%%%%%%%%%%%%%%%%%%%%%%%%%%%%%%%%%%%%%%%%%

The implicit occurence of the double-well GL kink in the Davydov
$\beta$ displacements offers the opportunity to introduce the multifractality
issue and closely related intermittent features,
which presumably are concepts of much interest for the dynamical phases of
biological chains.
Previously, Brax \cite{Br92} showed
the equivalence of the GL equation with real coefficients
under random initial conditions and the linear heat equation with
Gaussian random potential and provided a multifractal analysis of the
problem of direct relevance for our considerations. As it is well-known the GL
equation, which is a cubic
reaction- diffusion equation, describes phenomenologically the
evolution of the order parameter in superconductive phase transitions,
and models also spatial and time fluctuations of systems near
Hopf (oscillatory) bifurcations.
Moreover, in 1972, Scalapino, Sears and Ferrell \cite{Sca72} studied in
detail the statistical mechanics of one-dimensional GL
fields. They remarked that such fields can describe the dynamical
behaviour of nearly-ordered systems which are not undergoing sharp phase
transitions, and conjectured that the real-field case may have
application in some organic chain systems. Our arguments support that idea.
In the D'Alambert variable ($\xi=x-v_{s}t$), the GL kink $K_{GL}$ is
the solution of a GL equation of the type
%%%%%%%%%%%%%%%%%%%%%%%%%%%%%%%%%%%%%%%%%%%%%%%%%%%%
\begin{equation}\label{eq:r20}
\frac{\partial K_{GL}}{\partial t}
\left(\equiv - v_s\frac{d K_{GL}}{d \xi}\right )
=\frac{d ^{2} K_{GL}}{d \xi ^{2}}
+p K_{GL}-r K_{GL}^{3}
\end{equation}
%%%%%%%%%%%%%%%%%%%%%%%%%%%%%%%%%%%%%%%%%%%%%%%%%%%
with real $p$ and $r$ coefficients.
In the static case such GL equations are typical for the structural
phase-transitions \cite{Gor83} in equilibrium situations, but
they can be used also in nonequilibrium, slowly-driven systems.
Following Brax, see also \cite{Can93}, one can develop a multifractal (MF)
formalism, and also study the weak intermittent properties of the GL equation
(and also of the Davydov's model),
if the kink probabilistic distribution function is identified with
the $\tau$- function of the MF formalism
\begin{equation}\label{eq:r1}
{\cal G}(\xi _{2})- {\cal G}(\xi _{1}) \; \approx
\int _{\xi _{1}}^{\xi _{2}} \beta (\xi ')d\xi '\; \equiv -\tau (\xi;q)
\end{equation}
The $q$-parameter is well-known in the MF approach, being a
convenient way of describing the mathematical properties of the sets of
local growth probabilities of a physical quantity.
We recall that in the MF formalism the function $\tau (q)$ when it exists
characterizes the power-law scaling of the cumulant generating function
\cite{hal}.
The derivative of this function
with respect to $q$ is denoted by $\alpha$ and via the Lagrange
multipliers procedure of statistical thermodynamics, one obtains the
function $f(\alpha)$ which is the density of a measure, and can be
interpreted as a fractal dimension when it is positive, and related
to instabilities for negative values. The equations
$\alpha =\delta \tau /\delta q$ and $f(\alpha)=q\alpha -\tau$ represent the
basis
of the MF formalism, see e.g., \cite{Man89}. Formally, $q$ is
the inverse temperature, $\tau$ is the Gibbs free energy, and $f$ is the
entropy.
One can plot the second derivative of the $\tau$-function (the `specific
heat') and find out intervals of the `temperature' variable, within
which the plot clearly displays features of a phase transition, that
is a peak in the `specific heat' at a certain value of the
`temperature' variable \cite{Can93}.
At the same time, in our interpretation, $q$ is the exponent of the
local mean polarization of electric dipoles $P_i^q$ over sets of
self-similar dipolar clusters, which in fact is another
common assumption of the MF formalism. We hope to study in detail these
heuristic claims in future publications.

In conclusion, in our opinion, the old conjecture of Scalapino, Sears
and Ferrell \cite{Sca72} concerning incipient GL-type phase transitions in
one-dimensional organic chains might be at a quite good place within
the dynamical Davydov's model of protein chains. Moreover, one can attribute
self-organized features to the
non-equilibrium competition between the two dynamical phases as Canessa
and myself have shown in a previous paper \cite{Rc93}. Multifractal and
intermittent scaling in the present biological context is of great interest.

Regarding our conjecture of relating Davydov's model to dynamic-structural
transitions, it is to be also mentioned the debate on the connection
between the equations of motion for Davydov's solitons
and the $\phi ^4$-chain \cite{China}.

Finally, we recall that metastable, biological- `ferroelectric' states have
been first considered by Fr\"ohlich \cite{f73}.

%%%%%%%%%%%%%%%%%%%%%%%%%%%%%%%%%%%%%%%%%%%%%%%%%%%%%%%%%%%%%%%%%%%%%%%%%%%%
\section*{Acknowledgments}

The work was supported in part by the CONACyT Project 4868-E9406.
The author is grateful to Dr. E. Canessa for useful discussions .
%%%%%%%%%%%%%%%%%%%%%%%%%%%%%%%%%%%%%%%%%%%%%%%%%%%%%%%%%%%%%%%%%%%%%%%%%%%%%

%%%%%%%%%%%%%%%%%%%%%%%%%%%%%%%%%%%%%%%%%%%%%%%%%%%%%%%%%%%%%%%%%%%%%%%%%%%%

\end{document}